\documentclass[12pt]{article}
\usepackage{epsfig,graphicx}
\usepackage[totalwidth=170mm, totalheight=224mm]{geometry}

\newcommand{\beq}{\begin{equation}}
\newcommand{\eeq}{\end{equation}}
\newcommand{\bea}{\begin{eqnarray}}
\newcommand{\eea}{\end{eqnarray}}

\newcommand{\gsim}{\lower.7ex\hbox{$
\;\stackrel{\textstyle>}{\sim}\;$}}
\newcommand{\lsim}{\lower.7ex\hbox{$
\;\stackrel{\textstyle<}{\sim}\;$}}

\def\lsim{\mathrel{\rlap{\lower3pt\hbox{\hskip0pt$\sim$}}
    \raise1pt\hbox{$<$}}}         %less than or approx. symbol
\def\gsim{\mathrel{\rlap{\lower4pt\hbox{\hskip1pt$\sim$}}
    \raise1pt\hbox{$>$}}}         %greater than or approx. symbol

\newcommand{\bibit}[1]{\bibitem{#1}}

\newcommand{\aver}[1]{\langle #1\rangle}

\newcommand{\La}{\overline{\Lambda}}
\newcommand{\Si}{\overline{\Sigma}}
\newcommand{\Lam}{\Lambda_{\rm QCD}}

\newcommand{\GeV}{\,\mbox{GeV}}
\newcommand{\MeV}{\,\mbox{MeV}}
\newcommand{\matel}[3]{\langle #1|#2|#3\rangle}
\newcommand{\state}[1]{|#1\rangle}

\newcommand{\vep}{\varepsilon}

\newcommand{\msp}[1]{\mbox{\hspace*{#1mm}~}}

\begin{document}
\thispagestyle{empty}
\vspace*{-10mm}

\begin{flushright}
Bicocca-FT-03-28\\
UND-HEP-03-BIG\hspace*{.08em}02\\
hep-ph/0309081\\
\vspace*{2mm}
\end{flushright}
\vspace*{15mm}

\boldmath
\begin{center}
{\LARGE{\bf
Recent Advances in Semileptonic $B$ Decays }}
\vspace*{15mm} 

{\tt 
Invited talk given at FPCP\,2003; {\sf ~Paris, \,June 3-6,\, 2003}}

\vspace*{7mm} 
\end{center}

\unboldmath
\smallskip
\begin{center}
{\Large{Nikolai Uraltsev$^{\,*}$
}}  \\

\vspace{4mm}
{\sl INFN, Sezione di Milano, Milan, Italy} 
\vspace*{18mm}

{\large {\bf Abstract}}\vspace*{-.9mm}\\
\end{center}
\noindent
Aspects of the OPE-based QCD theory of $B$ decays are discussed. We
have at least one nontrivial precision check of the OPE at the
nonperturbative level in inclusive decays. The data suggest proximity
to the `BPS' limit for $B$ mesons. Its consequences are addressed and
their accuracy qualified. It is suggested that theory-wise
$B\!\to\!D\,\ell\nu$ near zero recoil offers an accurate alternative
way to measure $|V_{cb}|$.  It is shown that the OPE is in a good
standing when confronted by experiment. The alleged controversy
between theory and data on certain decay channels is found to be an
artefact of oversimplifying the OPE in presence of high experimental
cuts severely degrading the effective hardness. The effects
exponential in the latter are missed in the traditionally used
expressions, yet they do not signify a breakdown of the $\,1/m_b$
expansion proper. They typically increase the apparent $b$ quark mass
in $B\!\to\! X_s+\gamma$ by $70\MeV$ or more, together with an even
more dramatic downward shift in the kinetic expectation value. The
utility of the second moment of $\,E_\gamma\,$ is emphasized once the
aforementioned effects have been included. Incorporating them brings
different measurements into a good agreement, provided the OPE-based
theory employs the robust approach.

\setcounter{page}{0}

\vfill

~\hspace*{-12.5mm}\hrulefill \hspace*{-1.2mm} \\
\footnotesize{%% \noindent 
\hspace*{-5mm}$^*$On leave of absence from Department of 
Physics, University of Notre Dame, Notre Dame, 
IN 46556, USA \hspace*{-7mm}~\\
\hspace*{-5mm}\hspace*{.25em} and St.\,Petersburg Nuclear Physics 
Institute, Gatchina, St.\,Petersburg  188300, Russia}
\normalsize

\newpage

\section{Introduction}
The QCD-based theory of inclusive heavy flavor decays by now 
has made a long journey from its first fundamental results beginning 
with the QCD theorem \cite{qcdtheor} 
which established absence of the leading $\Lam/m_b$
power corrections to total decay rates. Based on the Wilsonian OPE, 
the heavy
quark expansion is 
now a well developed field of QCD with many nontrivial
phenomenological applications \cite{rev,ioffe}.
The new generation of data provides accurate measurements of
many inclusive characteristics in $B$ decays. In practical terms, \\
$\bullet$ An accurate and reliable determination of some heavy
quark parameters is available directly from experiment;\\
$\bullet$ Extracting $|V_{cb}|$ from $\Gamma_{\rm sl}(B)$ has good
accuracy and solid grounds;\\
$\bullet$ We have at least one nontrivial precision check of the OPE at
the nonperturbative level.\\ 
The latter comes comparing the average lepton energy with
the invariant hadronic mass, as explained in Sect.~3.

The present theory allows to aim for a percent accuracy in $|V_{cb}|$.
Such a precision becomes possible owing to a number of theoretical
refinements. 
The low-scale running masses $m_b(\mu)$, $m_c(\mu)$, the
expectation values $\mu_\pi^2(\mu)$, $\mu_G^2(\mu)$, ... are completely
defined and can be determined from experiment with an in principle
unlimited accuracy. Violation of local duality potentially
limiting theoretical predictability, has been scrutinized and found 
to be negligibly small in total semileptonic
$B$ widths \cite{vadem}; it can be controlled experimentally with
dedicated high-statistics measurements.
Present-day perturbative technology makes computing  $\alpha_s$-corrections
to the Wilson coefficients of nonperturbative operators feasible.
It is also understood how to treat higher-order power
corrections in a way which renders them suppressed \cite{imprec}.

The original motivation for the precision control of strong
interaction effects in $B$ decays was the quest for 
$|V_{cb}|$ and $|V_{ub}|$ as means to cross check the Standard
Model. Interesting physics
lies, however not only in the CKM matrix; knowledge of heavy quark masses and
nonperturbative parameters of QCD is of high importance as well.

Some of the heavy quark parameters 
like $\mu_G^2$ are known beforehand. Proper 
field-theoretic definition allows its accurate determination 
from the $B^*\!-\!B$ mass splitting:
$\mu_G^2(1\GeV)\!=\!0.35^{+.03}_{-.02}\GeV^2$ \cite{chrom}. 
A priori less certain is $\mu_\pi^2$. 
However, the inequality $\mu_\pi^2\!>\!\mu_G^2$ valid 
for any definition of
kinetic and chromomagnetic operators respecting the QCD commutation
relation $[D_j,D_k]\!=\!-ig_sG_{jk}$, 
and the corresponding sum rules 
essentially limit its range: $\mu_\pi^2(1\GeV)\!=\!0.45\!\pm\!
0.1\GeV^2$.

The important feature of the heavy quark parameters -- in particular
the nonperturbative ones -- is that they represent well-defined physical
quantities which take universal values independent of a process under
consideration. As such they can be determined separately in quite
different processes with $b$ quarks. The best example is $m_b$ itself
-- it can be measured through the near-threshold production $e^+e^-
\to b\bar{b}$, or in inclusive $B$ decays. The extracted value must
be the same, for it is just an actual $b$ quark mass in actual QCD.

Recent experimental data on inclusive $B$ decays generally show a 
nontrivial agreement between
quite different -- and {\it a priori}\, unrelated -- measurements at
the nonperturbative level, on one hand, and consistency with the
QCD-based OPE theory. It is fair to note that some
discrepancies have been reported as constituting obvious
problems for the OPE. 
In my opinion, however, these claims are unjustified and 
rather rooted in a sloppy application of the heavy quark expansion,
in particular of the OPE.  The proper treatment, in fact seems to yield
a good agreement between different measurements, well within correctly
assessed theoretical accuracy. I shall illustrate this later 
in Sect.~3. At this point it is interesting to note that all data
seem to point at a relatively low kinetic expectation value $\mu_\pi^2
\lsim 0.4\GeV^2$; the individual uncertainties are sizable, though.

\section{BPS\, limit}

An intriguing theoretical environment opens up if
$\mu_\pi^2(1\GeV)$ 
is eventually confirmed to be close enough to
$\mu_G^2(1\GeV)$ as currently suggested by experiment, say it does
not exceed $0.45\GeV^2$. If $\mu_\pi^2\!-\!\mu_G^2
\!\ll \!\mu_\pi^2$ it is advantageous to analyze strong dynamics 
expanding around the point
$\mu_\pi^2\!=\!\mu_G^2$ \cite{chrom}. This is not just 
one point of a continuum in
the parameter space, but a quite special `BPS' limit where the
heavy flavor  
ground state satisfies functional relations 
$\vec{\sigma}\vec{\pi}\state{B}\!=\!0$. This limit is 
remarkable in
many respects; for example, it saturates the bound \cite{newsr}
$\varrho^2\!\ge\!\frac{3}{4}$ for the slope of the IW function. 
It should be 
recalled that already quite some time ago there were
dedicated QCD sum rules estimates of both $\varrho^2$
\cite{bshifmanrho} and $\mu_\pi^2$ \cite{lubl} yielding literal values 
nearly at the respective lower bounds, supporting this limit.

The SV heavy quark sum rules  
place a number of important constraints on the nonperturbative
parameters. For instance, they yield a bound on the IW slope
\beq
\mu_\pi^2\!-\!\mu_G^2 \!=\!3\tilde\vep^2
(\varrho^2\!-\!\mbox{$\frac{3}{4}$})\,, \qquad 0.4\GeV \lsim
\tilde\vep \lsim 1\GeV 
\label{56}
\eeq
so that $\varrho^2$ can barely reach $1$ being rather closer to
$0.85$. It is interesting that this prediction \cite{chrom} turned out
in a good agreement with the recent lattice calculation \cite{rholat}
$\varrho^2=0.83^{+.15+.24}_{-.11-.01}$. It 
leaves only a small window for the slope of the actual $B\!\to\! D^*$
formfactor, excluding values of $\hat\varrho^2$ 
in excess of $1.15\!-\!1.2$. This would be a
very constraining result for a number of experimental studies, in
particular for extrapolating the $B\to D^*$ rate to zero recoil. Since
there is a strong correlation between the extrapolated rate and the
slope, this may change the extracted value of $|V_{cb}|$. Therefore, it
is advantageous to analyze the $B\to D^* \ell\nu$ data including the
above constraint as an option, and I suggest our experimental
colleagues explore this in future analyses.

The
experimentally measured slope $\hat\rho^2$ differs from $\varrho^2$ by
heavy quark symmetry-violating corrections. The estimate by Neubert 
that  $\hat\rho^2$ is smaller than  $\varrho^2$, $\hat\rho^2\simeq
\varrho^2\!-\!0.09$ seems to be ruled out by experiment. It is not
clear if a better estimate can be made in a trustworthy way.

The whole set of the heavy quark sum rules is even more
interesting. Their constraining power depends strongly on the actual
value of $\mu_\pi^2$. When it is at the lower end of the allowed
interval, the BPS expansion appears the most effective way to analyze
all the relations. 
%%% \vspace*{2mm}

\subsection{Miracles of the BPS limit}
 
A number of useful
relation for nonperturbative parameters hold in the limit 
$\mu_\pi^2\!=\!\mu_G^2$: they include $\varrho^2\!=\!\frac{3}{4}$,
$\La\!=\!2\Si$, $\:\rho_{LS}^3\!=\!-\rho_D^3$, relations for nonlocal
corre\-lators $\rho_{\pi G}^3\!=\!-2\rho_{\pi \pi}^3$, $\rho_A^3\!+\rho_{\pi
G}^3\!=\! -(\rho_{\pi \pi}^3\!+\rho_{S}^3)$, etc.

This limit also extends a number of the heavy
flavor symmetry relations for the ground-state mesons 
to all orders in $1/m\,$:\\
$\bullet$ There are no formal power corrections to the relation
$M_P\!=\!m_Q+\La$ and, therefore to $m_b\!-\!m_c=M_B\!-\!M_D$. The
routinely used spin-averaged mass difference, however is not stable.\\
$\bullet$ For the $\;B\!\to\! D\;$ amplitude the heavy quark limit 
relation between
the two formfactors 
\beq
f_-(q^2)=-\frac{M_B\!-\!M_D}{M_B+M_D}\; f_+(q^2)
\label{106}
\eeq
does not receive power corrections.\\
$\bullet$ For the zero-recoil $\;B\!\to\! D\;$ amplitude all
$\,\delta_{1/m^k}\,$ terms vanish.\\
$\bullet$ For the zero-recoil formfactor $\,f_+\,$ controlling decays 
with massless leptons
\beq
f_+((M_B\!-\!M_D)^2)=\frac{M_B+M_D}{2\sqrt{M_B M_D}}
\label{108}
\eeq
holds to all orders in $1/m_Q$.\\
$\bullet$ At arbitrary velocity power corrections in $\;B\!\to\! D\;$
vanish,
\beq
f_+(q^2)=\frac{M_B+M_D}{2\sqrt{M_B M_D}} \;\,\mbox{{\large$ 
\xi$}}\!\left(\mbox{$\frac{M_B^2+M_D^2-
\raisebox{.6pt}{\mbox{{\normalsize $q^2$}}}}{2M_BM_D}$}\right)
\label{110}
\eeq
so that the $\;B\!\to\! D\;$ decay rate directly yields 
Isgur-Wise function $\xi(w)$.\\
It is interesting that experimentally the slope of the  $B\!\to\!
D\,$ amplitude is 
indeed smaller centering around $\hat\rho_{(D)}^2\!\approx \!1.15$ 
\cite{ckmrep}, indicating qualitative agreement with the BPS regime.

What about the $\;B\!\to\! D^*\,$ amplitude, are the corrections
suppressed as well? Unfortunately, the answer is negative. The
structure of power corrections indeed simplifies in the BPS limit,
however $\delta_{1/m^2}$,  $\delta_{1/m^3}$ are still very
significant \cite{chrom}, and the literal estimate for $F_{D^*}(0)$
falls even below $0.9$. Likewise, we expect too significant
corrections to the shape of the $B\!\to\! D^*$ formfactors. Heavy
quark {\tt spin} symmetry controlling these transitions
seems to be violently affected by strong interactions for charm.

A physical clarification must be added at this point. Absence of all
power corrections in $1/m_Q$ for certain relations may be naively
interpreted as implying that they would hold for arbitrary, even small
quark masses, say in $B\!\to\!K$ transitions. This is not correct,
though, for the statement refers only to a particular fixed order in
$1/m_Q$ expansion in the strict BPS limit. In fact the relations
become more and more accurate approaching this limit only above 
a certain  
mass scale of order $\La$, while below it their violation is of order
unity regardless of proximity of the heavy quark ground state to BPS.

\subsection{Quantifying deviations from BPS} 

Since the BPS limit
cannot be exact in actual QCD, it is important to understand the 
accuracy of its predictions.
The dimensionless parameter describing the deviation from BPS is
\beq
\beta =\|\pi_0^{-1}(\vec\sigma\vec\pi)\,\state{B}\| 
\equiv
\mbox{$\sqrt{3\!\left(\varrho^{2}\!-\!\frac{3}{4}\right)} =
3
\left(\sum \rule{0mm}{3mm} \right.
\mbox{\hspace*{-3.5mm}\raisebox{-2.2mm}{{\tiny $n$}}\hspace*{.35mm}}
 \left.\,|\tau_{1/2}^{(n)}|^{2}\right)^{\frac{1}{2}}$}.
\label{112}
\eeq
Numerically $\,\beta\,$ is not too small, similar in size to generic
$\,1/m_{c}\,$ expansion parameter, and the relations violated to
order $\beta$ may in practice be more of a qualitative nature only.
However, the expansion parameters like 
$\mu_\pi^2\!-\!\mu_G^2 \propto \beta^2$ can be good enough. Moreover,
we can count together powers of $1/m_c$ and 
$\beta$ to judge the real quality of a particular heavy quark relation.
Therefore understanding at which order in $\beta$ the BPS
relations get corrections is required. In fact, we need 
classification in powers
of $\beta$  to {\tt all} orders in $1/m_Q$.

Relations (\ref{106}) and (\ref{110}) for the $B\!\to\!D$
amplitudes at arbitrary velocity can get first order corrections in
$\beta$. Thus they may be not very accurate. The same refers to
equality of $\rho_{\pi G}^3$ and $-2\rho_{\pi\pi}^3$. The other
relations mentioned for heavy quark parameter are accurate up to 
order $\beta^2$.
The other important BPS relations hold up to order $\beta^2$ as well:\\
$\bullet$ $M_B\!-\!M_D=m_b\!-\!m_c$ and $M_D=m_c\!+\!\La$ \\
$\bullet$ Zero recoil matrix element $\matel{D}{\bar{c}\gamma_0 b}{B}$
is unity up to ${\cal O}(\beta^2)$\\
$\bullet$ Experimentally measured $B\!\to\!D$ formfactor $f_+$ near
zero recoil receives only second-order corrections in $\beta$ to all
orders in $1/m_Q$:
\beq
f_+\left((M_B\!-\!M_D)^2\right) = \frac{M_B\!+\!M_D}{2\sqrt{M_BM_D}} \;\,
+ {\cal O}(\beta^2)\;.
\label{116}
\eeq
This is an analogue of the Ademollo-Gatto theorem for the BPS
expansion. 

The similar statement then applies to $f_-$ as well, and the heavy 
quark limit
prediction for $f_-/f_+$ must be quite accurate near zero recoil. It can be
experimentally checked in the decays $B\!\to\!D\,\tau\nu_\tau$.

As a practical application of the results based on the BPS expansion, 
one can calculate the $B\!\to\!D$ decay amplitude near zero recoil to
use this channel for the model-independent extraction of $|V_{cb}|$ in
future high-luminosity experiments. For power corrections we
have 
\beq
\frac{M_B\!+\!M_D}{2\sqrt{M_BM_D}}\; f_+\left((M_B\!-\!M_D)^2\right) \;=\; 1
\; +
\left(
\frac{\La}{2} \!-\!\overline\Sigma\right) 
\left(\frac{1}{m_c}\!-\!\frac{1}{m_b}\right)
\frac{{M_B\!-\!M_D}}{M_B+M_D}-
{\cal O}\left(
\frac{1}{m_Q^2}\!\right)\; .
\label{124}
\eeq
We see that this indeed is of the second order in $\beta$. Moreover, 
$\La\!-\!2\Si\,$ is well constrained through $\mu_\pi^2\!-\!\mu_G^2$ by
spin sum rules. Including perturbative corrections (which should be
calculated  in the proper renormalization scheme respecting BPS regime) 
we arrive at the estimate 
\beq
\frac{2\sqrt{M_B M_D}}{M_B+M_D}\; f_+(0) = 1.03\pm 0.025 \msp{10}
\label{126}
\eeq
It is valid through order $\beta^2\frac{1}{m_c}\,$ accounting for \,{\tt all
powers}\, of $\,1/m_Q\,$ to order $\beta^1$. Assuming the counting rules
suggested above this corresponds to the precision up to $1/m_Q^4$,
essentially better than the ``gold-plated'' $B\!\to\!D^*$  formfactor where
already $1/m_Q^2$ terms are large and not well known. Therefore, the
estimate (\ref{126}) must be quite accurate. In fact, the major source
of the uncertainty seems to be perturbative corrections, which can be
refined in the straightforward way compared to decade-old
calculations.

\section{\boldmath Inclusive $B$ decays, OPE and heavy quark parameters}

Inclusive distributions
in semileptonic and radiative decays are the portal to accurately
determining the nonperturbative heavy quark parameters controlling
short-distance observables in $B$ decays. In this way one can extract
the most precise and model-independent value of $V_{cb}$ from
$\Gamma_{\rm sl}$, and possibly of $V_{ub}$ if a number of
experimental problems can be overcome.

High accuracy can be achieved in a comprehensive approach
where many observables are measured in $B$ decays to extract necessary
`theoretical' input parameters. It is crucial that here one can do
without relying on charm mass expansion at all, i.e.\ do not assume
charm quark to be really heavy in strong interaction mass scale. 
For reliability of the $1/m_c$ expansion is questionable. 
Already in the $1/m_Q^2$ terms one has $\frac{1}{m_c^2}
\!>\!14\frac{1}{m_b^2}$; even for the worst mass scale in the 
width expansion,   
$\frac{1}{(m_b\!-\!m_c)^2}$ is at least $8$ times smaller 
than $\frac{1}{m_c^2}$. 
There are indications \cite{chrom} that the 
nonlocal correlators
affecting meson masses can be particularly large -- a
pattern independently observed in the 't~Hooft model \cite{lebur}. This
expectation is supported by the pilot lattice study \cite{kronsim2}
which -- if taken at face value -- 
suggests a very large value of a particular combination 
$\rho_{\pi\pi}^{3\!}\!\!+\!\rho_{S}^{3}$ entering in the 
conventional approach. On the other hand, 
non-local correlators are not measured in inclusive $B$ decays, so
that assumptions about them can hardly be verified. 

The approach which is free from relying on charm mass
expansion \cite{amst} was put forward at the CKM-2002 Workshop at
CERN. It allows to utilize the full 
power of the comprehensive studies, and  makes use of 
a few key facts \cite{optical,motion}:\\
~\hspace*{.2em}$\bullet$ Total width to order $1/m_b^3$ is affected by
a single new Darwin operator; 
the moments depend also on $\rho_{LS\,}^3$, albeit weakly.\\
~\hspace*{.2em}$\bullet$ No nonlocal correlators ever enter {\it per se}.\\
~\hspace*{.2em}$\bullet$ Deviations from 
the HQ limit in the expectation values are
driven by the maximal mass scale, $2m_b$ (and are additionally
suppressed by proximity to the BPS limit); they are negligible in
practice. \\
~\hspace*{.2em}$\bullet$ Exact sum rules and inequalities  
hold for properly defined Wilsonian parameters.
\vspace*{1mm}

The strategy and how it works in practice has been described in detail in
Refs.~\cite{amst,ckm03}. Here I only recall a few salient facts. 

The gross features of the lepton energy distribution 
are mainly shaped by the parton
expressions and get only slightly corrected by nonperturbative
effects. Consequently, a few lowest lepton moments depend primarily on
one and the same combination of the heavy quark masses, 
$m_b \!-\! 0.7 m_c$, with insufficient sensitivity to the nonperturbative
operators. This does not allow in practice to constrain more than one
combination of the quark masses. Yet even these moments turn out
informative for $V_{cb}$, since $\Gamma_{\rm sl}$ depends on nearly the
same combination of $m_b$ and $m_c$.

Moments of invariant mass square $M_X^2$ in the final state of 
semileptonic $B$
decays appear more important in constraining the nonperturbative
parameters, since it is {\it ab initio}\, dominated by nonperturbative 
terms. However, the first hadronic moment $\aver{M_X^2}$ turns out
to depend on practically the same combination
$m_b\!-\!0.7m_{c\!}+\!0.1\mu_\pi^2\!-\!0.2\rho_D^3$ as the lepton
moment. (Here and below all coefficients are assumed to be in the
corresponding powers of GeV.)
Not very constraining, this provides, however a highly nontrivial check
of the HQ expansion. For example, taking DELPHI's
central value for
$\aver{M_X^2}$ we would predict $\aver{E_\ell}\!=\!1.375\GeV$, while
experimentally they obtain $\aver{E_\ell}\!=\!(1.383\pm 0.015)\GeV$ 
\cite{newdelphi}.
In this respect such a comparison is more critical than among the 
lepton moments themselves. In particular, these two first 
moments together verify the heavy
quark sum rule for $M_B\!-\!m_b$ with the accuracy about $40\MeV$!

The higher hadronic moments (with respect to the average) are already 
more directly sensitive to nonperturbative parameters $\mu_\pi^2$ and
$\rho_D^3$. 
Ideally, they would measure the kinetic and Darwin expectation values
separately. At present, however, we have only an approximate
evaluation and informative upper bound on $\tilde\rho_D^3$. The
current sensitivity to $\mu_\pi^2$ and $\tilde\rho_D^3$ is about
$0.1\GeV^2$ and $0.1\GeV^3$, respectively. A more accurate measurement
would also require refined theoretical description which is possible
once the modified higher hadronic 
moments are measured at $B$-factories \cite{ckm03}. 

The experimental constraints on the combination driving 
$\Gamma_{\rm sl}(B)$ appear stronger for 
the hadronic moment. Using it instead of  $\aver{E_\ell}$ we would
arrive at 
{\small
$$
\mbox{{\large
$\frac{|V_{cb}|}{0.042}$}} 
\!=\! 
%%%\mbox{\hspace*{-.08mm}~}\raisebox{-.2mm}{\mbox{{\normalsize$1$}}}
1\,+\, 0.14\,
[\aver{M_X^2}\!-\!4.54\GeV^2] \,-\, 0.03\, (m_c\!-\!1.15\GeV) 
+\, 0.1\,(\mu_\pi^2\!-\!0.4\GeV^2)
 +0.1\,(\tilde\rho_D^3\!-\!0.12\GeV^3)\,.
\label{45}
$$
}
One finds that measuring the second and third hadronic moments which
directly constrain  $\mu_\pi^2$ and $\tilde\rho_D^3$ is 
an essential step in implementing the comprehensive program of extracting
$|V_{cb}|$ (see figures in \cite{DELPHI,newdelphi}). 
To illustrate, neglecting possible theoretical uncertainties in
the above relations, we get
\beq
|V_{cb}|=0.0418\,\left(1\pm 0.01_{\rm SL\,width}  
\pm 0.02_{\rm HQ\,par}\right)
\label{46}
\eeq
from only the DELPHI hadronic moments.
It is crucial that
this extraction carries no hidden assumptions, and at no point we
rely on $1/m_c$ expansion. Charm quark could be either heavy, or
light as strange or up quark, without deteriorating -- and rather
improving -- the accuracy.

Incorporating other measurements further decreases the experimental 
uncertainty associated
with the heavy quark parameters. However, another limiting factor
becomes important -- the accuracy of the theoretical expression for
the moments used in the analyses. All the theoretical expressions relied 
upon have a limited accuracy, a fact discarded above for simplicity.

A similar analysis can be applied to the moments with a cut on lepton
energy; in particular, CLEO measured a few lepton energy moments for
$E_\ell \!>\! 1.5\GeV$ with unprecedented accuracy. They
also fix more or less the same combination of masses and
nonperturbative parameters. The value comes out close, but does not
literally coincide with that obtained by DELPHI. This sometimes is referred
to as a problem for theory. My opinion is different -- we do not see
here a convincing evidence of the theory failure: the intrinsic theory
accuracy is essentially degraded by the relatively high cut employed,
with the effective hardness deteriorated. The additional 
theoretical uncertainties are typically overlooked here.

\subsection{Cuts and extracting heavy quark parameters}  

In order to enjoy in full the potential of a small expansion
parameter provided by the heavy quark mass, 
the observable in question must be sufficiently inclusive. However
experimental cuts imposed for practical reasons -- to suppress
backgrounds etc. -- often essentially degrade the effective hardness
${\cal Q}$ of the process.  This brings in another expansion
parameter $1/{\cal Q}$ effectively replacing $1/m_b$ in certain QCD
effects. The reliability of the expansion greatly deteriorates for
${\cal Q}\!\ll\! m_b$.  This phenomenon is particularly important in
$b\!\to\! s+\gamma$ decays where experiments so far have imposed cut
$E_\gamma\! >\!2\GeV$ or even higher.

The theoretical aspects of such limitations have been discussed
during the last couple of years \cite{uses,amst,ckm03}. In
particular, the effective `hardness' of the inclusive 
$b\!\to\! s+\gamma$ decays with $E_\gamma\! >\! 2\GeV$ amounts to only about
$1.25\GeV$, which casts doubts on the precision of the routinely used
expressions incorporated into the fits of heavy quark parameters.

Ref.~\cite{misuse} has analyzed the numerical aspects and it was shown
that these effects turn out 
significant, lead to a systematic bias that often exceeds naive 
error estimates and therefore cannot be ignored. 
Evaluating them in the most straightforward
(although somewhat simplified) way we find, for instance for $b\to
s+\gamma$ decays 
\bea
\nonumber
\tilde m_b  & \msp{-3}\simeq \msp{-3} & m_b+ 70\MeV \\
\tilde \mu_\pi^2  & \msp{-3}\simeq \msp{-3} &  
\mu_\pi^2 - (0.15\div 0.2)\GeV^2
\label{12}
\eea
where $\tilde m_b$ and $\tilde \mu_\pi^2\,$ are the {\tt apparent}
values of the $b$ quark mass and of the kinetic expectation value,
respectively, as extracted from the $b\!\to\! s+\gamma$ 
spectrum in a usual way with $E_{\gamma} \!>\! 2\GeV$.  
Correcting for this bias eliminates
alleged problems for the OPE in describing different data and rather
leads to a too good agreement between the data on different types of
inclusive decays. 

Moreover, this resolves the controversy 
noted previously: while the values of $\La$ and $\mu_\pi^2$
reportedly extracted from the CLEO \,$b\!\to\! s+\gamma$ spectrum were
found to be significantly below the theoretical expectations, the
theoretically obtained spectrum itself turned out to yield a good 
description of the observed spectrum when we evaluated it based on
these theoretically preferred values of parameters \cite{uses}. 

The origin of these effects and why they are routinely missed in 
the standard application of the OPE and in estimates
of the theoretical accuracy, have been discussed in detail in 
Ref.~\cite{ckm03}. Conceptually this is related to the limited range 
of convergence of
the OPE for the width, determined in this case by the support of the
heavy quark distribution function. Referring to the original
publications \cite{ckm03,misuse} for details, here I give only the
numerical estimates for the cut-induced bias in $b\!\to\! s+\gamma$
decays. Namely, we can compare
the {\tt apparent} values of the $b$ quark mass $\tilde m_b$ and of the
kinetic operator $\tilde \mu_\pi^2$ with the true ones $m_b$,
$\mu_\pi^2$ which would be measured if no cut were 
imposed, Fig.~1. The deficit in $\La$ turns out quite significant. 

\begin{figure}[htb]
\begin{center}
\epsfig{file=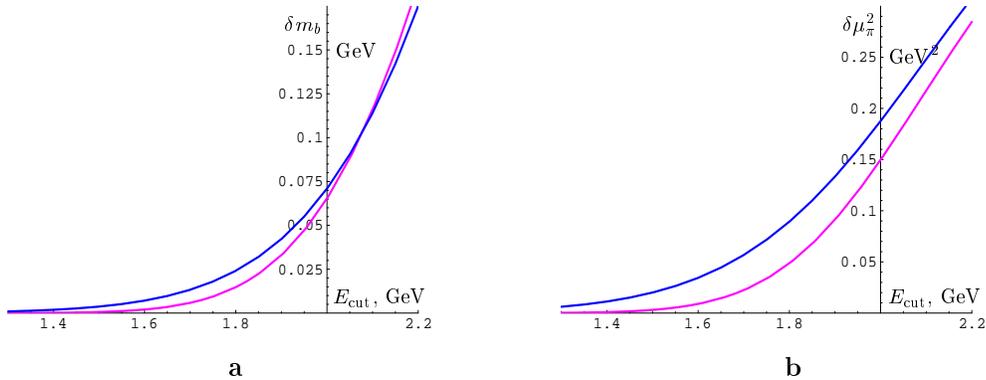,width=130mm}\vspace*{-3mm}
\caption{{\small 
\,The shifts $\tilde m_b\!-\!m_b$ in the quark mass \,({\bf a})\, and
$\mu_\pi^2\!-\!\tilde\mu_\pi^2$ in the kinetic operator \,({\bf b})\, 
introduced by imposing a lower cut in the photon
energy in $B\!\to\! X_s\!+\!\gamma$. Blue and maroon curves
correspond to two different ans\"{a}tze for the heavy quark distribution
function.\vspace*{-5mm}
}}
\label{fig:uraltsev-fig1}
\end{center}
\end{figure}

The naive extraction of the kinetic expectation value through the
variance of the truncated distribution undercounts it even more
dramatically as Fig.~1b illustrates, since higher moments are more
sensitive to the tail of the distribution.

Why are the effects of the cut so large? They are exponential in
the inverse hadronic scale $\mu_{\rm hadr}$, but are governed by the 
{\tt hardness} ${\cal Q}\!\simeq\! m_b\!-\!2E_{\rm cut}$ rather than by $m_b$: 
\beq
\La-\tilde\Lambda(E_{\rm cut}) \;\propto\; \mu_{\rm hadr}\;
e^{-\frac{{\cal Q}}{\mu_{\rm hadr}}}\;, \msp{10}
\mu_\pi^2-\tilde \mu_\pi^2 \; \propto \;\mu_{\rm hadr}^2\;
e^{-\frac{{\cal Q}}{\mu_{\rm hadr}}}
\label{38}
\eeq
(the exponent may have a form of a power of ${\cal Q}/\mu_{\rm
  hadr}$). Even at $m_b\to \infty$ these effects survive unless
${\cal Q}$\, is made large as well!
\vspace*{2mm}

From this comparison we conclude:\vspace*{.5mm}\\
$\bullet$ the value of $m_b$ as routinely extracted from the $b\!\to\!
s+\gamma$ spectrum is to be decreased by an amount of order
$70\MeV$;\vspace*{.5mm}\\
$\bullet$ relative corrections to $\mu_\pi^2$ are even more
significant and can naturally constitute a shift of
$\,0.2\GeV^2$. This arises on top of other potential effects.
%%% \vspace*{2mm}

%%% \noindent
\subsection{Practical implications}

One can imagine a dedicated analysis of the consistency of the OPE
predictions for inclusive $B$ decays through a
simultaneous fit of all available experimental data with the
underlying heavy quark parameters. For illustrative purposes here I adopt
instead a much simpler and transparent procedure 
which yet gives convincing
evidence that the OPE is in a good shape, and elucidates the actual
root of alleged problems. Let me assume a rather arbitrary 
choice $m_b\!=\!4.595\GeV$, $m_c\!=\!1.15\GeV$,
$\mu_\pi^2\!=\!0.45\GeV^2$, $\tilde\rho_D^3\!=\!0.06\GeV^3$ and 
$\rho_{LS}^3\!=\!-0.15\GeV^3$, natural from the theoretical
viewpoint. The precise values of the quark masses are adjusted, 
however to literally 
accommodate $\aver{M_X^2}_{E_\ell>1\,{\rm GeV}}\,$ now well measured
by BaBar and CLEO. We then obtain 
\bea
\nonumber
\rule{0mm}{5mm}\aver{M_X^2} &\msp{-5}\simeq\msp{-5} & 
4.434\GeV^2 \msp{5}[\mbox{cf. } (4.542\pm 0.105) \GeV 
\mbox{ (DELPHI)}\,] \\
\nonumber
\rule{0mm}{5mm}\aver{M_X^2}_{E_\ell>1.5\,{\rm GeV}} &\msp{-5}\simeq\msp{-5} & 
4.177\GeV^2 \msp{5.1}[\mbox {cf. } 4.180\GeV^2 \mbox{ (BaBar), } 
4.189\GeV^2 \mbox{ (CLEO)}\,] \\
\nonumber
\rule{0mm}{5mm}\aver{E_\ell} &\msp{-5}\simeq\msp{-5} &  1.389\GeV
\msp{6.9}[\mbox {cf. } (1.383 \pm 0.015)\GeV \mbox{ (DELPHI)}\,] \\
\nonumber
\rule{0mm}{5mm}\aver{E_\gamma}_{E_\gamma>2\,{\rm GeV}} 
&\msp{-5}\simeq\msp{-5} & 
2.329\GeV \msp{6.9}[\mbox {cf. } (2.346 \pm 0.034)\GeV  \mbox{ (CLEO)}\,]\\ 
\rule{0mm}{5mm}
\aver{E_\gamma^2\!-\!\bar E_\gamma^2}_{E_\gamma>2\,{\rm GeV}} 
&\msp{-5}\simeq\msp{-5} &\msp{-.9}
^{0.0202}_{0.0233} \,\GeV^2 \msp{5.0}
[\mbox {cf. } (0.0226 \pm 0.0066\pm0.0020)\GeV^2  \mbox{ (CLEO)}\,] \qquad
\label{56mis}
\eea
The moments of the photon spectra have been calculated including the
perturbative corrections as in Ref.~\cite{misuse}, i.e.\ they incorporate
the above `exponential' cut-related shifts at 
face value.\footnote{The two values for the second $E_\gamma$-moment
correspond to different light-cone ans\" atze; they are 
obtained discarding higher-order
power corrections to the light-cone distribution function.} 
Their counterpart for the semileptonic decays has not been 
included here, however. I also
show in Fig.~2 \,the $E^\ell_{\rm cut}$-dependence of $\aver{M_X^2}$
as it comes
from the literal OPE expressions with the above values of parameters.
Experimental data are from 
Refs.~\cite{newdelphi,newbabar,newcleo,cleobsg}.
%%% Refs.~\cite{newdelphi - cleobsg}.

\begin{figure}[htb]
\begin{center}
\epsfig{file=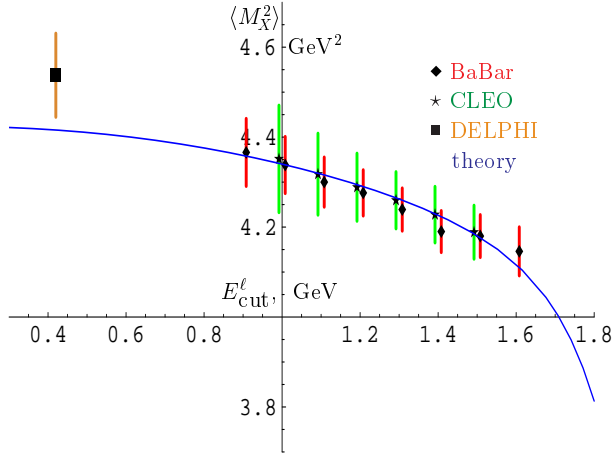,width=80mm}\vspace*{-4mm}
\caption{{\small 
Experimental values of the average hadronic mass square
$\aver{M_X^2}$ in $B\!\to\! X_c\,\ell\nu$ at different lower cuts in
$E_\ell$  and  the literal OPE prediction (blue curve) for the
stated heavy quark parameters. 
The DELPHI point assumes no cut on $E_\ell$. \vspace*{-5mm}
}}
\label{fig:uraltsev-fig2}
\end{center}
\end{figure}

A quick glance shows that there is hardly any disagreement with
theory, for both semileptonic and $b \!\to\! s+\gamma$ channels. Moreover,
even the $E^\ell_{\rm cut}$-dependence is naturally reproduced. This
radically differs from the results reported in 
\cite{gfit}. I believe that what we have heard from Ligeti in this
respect at this
conference about the alleged problems for the OPE in describing the
data, is not true.

As mentioned above, the first leptonic moment
$\aver{E_\ell}$ with cut at $1.5\GeV$, the CLEO's $R_1$ comes out slightly
different: $1.776\GeV$ vs.\ $1.7810\GeV$ measured by CLEO. The
difference is certainly significant if viewed experimentally. However,
is it really, if examined by theory?

%%% \noindent
\subsubsection{Cuts in semileptonic moments}

It was argued in Ref.~\cite{ckm03}\, that similar cut-related `exponential'
biases missed in the naive OPE applications affect the truncated
moments in the semileptonic decays as well. Their description, even 
simplified is less transparent and would be more involved,
though, being given by a more complicated and non-universal object
rather than the light-cone heavy quark distribution. Although the
numerical aspects are rather uncertain, it is understood that they can
be significant: the effective hardness 
\beq
{\cal Q}_{\rm sl}\simeq m_b-E_{\rm cut}-\sqrt{E_{\rm cut}^2+m_c^2} 
\label{80}
\eeq
at $E_{\rm cut}\!=\!1.5\GeV$ is about $1.25\GeV$ \cite{amst}, nearly 
the same as for $B\!\to\! X_s+\gamma$\, with $E_\gamma \!\gsim\! 2\GeV$.
Ref.~\cite{misuse} estimated that the 
exponential terms in semileptonic decays with 
$E_{\rm cut}\!\simeq\! 1.5\GeV$ can introduce effects of the same scale as
shifting $m_b$ upward by up to $25$ to $30\MeV$ (assuming fixed
$m_c$ and other heavy quark parameters). This rule of thumb --
although very tentative -- is
useful to get an idea of the ultimate theoretical accuracy one can
count on here. 

For example, the CLEO's cut moment 
$\,R_1\!=\!\aver{E_\ell}\raisebox{-1.3mm}{\mbox{{\tiny 
$\!E_\ell\!\! >\!\!1.5$GeV\,}}}$ is approximately given by \cite{amst}
\beq
R_1\!=\!1.776\GeV + 0.27(m_b\!-\!4.595\GeV) -
0.17(m_c\!-\!1.15\GeV) 
\msp{7}\mbox{at }
 |V_{ub}/V_{cb}|\!=\!0.08
\label{94}
\eeq
(the above mentioned values of the nonperturbative parameters are
assumed). An increase in $m_b$ by only $20 \MeV$ would then change
\beq
R_1 \to R_1+ 0.0055\GeV
\label{96}
\eeq
%%% and would 
perfectly fitting the central CLEO's value
$1.7810\GeV$. It is worth noting that the above equations 
make it evident that the imposed cut on $E_\ell$ degrades 
theoretical calculability of $R_1$ far beyond its experimental error
bars, the fact repeatedly emphasized over the last
year. Unfortunately, this was not reflected in the fits of
parameters which placed
much weight on the values of $\,R_0\,$--$\,R_2\,$ just owing to their
small experimental uncertainties, whilst paying less attention to actual
theoretical errors.

Even stronger reservations should be applied to the theoretical
predictability for CLEO's $\,R_2$
representing the second moment with the cut, keeping in mind 
that the effective hardness deteriorates for higher
moments. 

The CLEO's ratio $R_0$ is the normalized decay rate with the cut on
$E_\ell$ as high as $1.7\GeV$, and for it hardness ${\cal Q}$ is
below $1\GeV$. A precision -- beyond just semiquantitative -- treatment
of nonperturbative effects is then questionable, and far more significant
corrections should be allowed for. 
\vspace*{2mm}

We see that -- contrary to what one could typically hear over the last
year about experimental check of dynamical heavy quark theory  --
there is a good agreement of most data referring to
sufficiently `hard' decay distributions with the theory based on the
OPE in QCD, if the `robust' OPE approach is used. The latter was put forward
\cite{amst} to get rid of unnecessary vulnerable assumptions 
of usually employed fits of the data.
This consistency likewise refers to the absolute values of the 
heavy quark parameters
necessary to accommodate the data. Theory itself reveals, however that
the expansion becomes deceptive with an increase in the experimental
cuts. Here I mentioned the clearest effects, those from the
variety of `exponential' terms in the effective hardness. While
presently not amenable to precise theoretical treatment, they can
be estimated using most natural assumptions and are found to be very
significant for $E^{\ell}_{\rm cut}\!\gsim\! 1.5\GeV$ 
and $E^{\gamma}_{\rm cut}\!\gsim\! 2\GeV$ often employed in
experiment. Taking these corrections at face value and incorporating
in our predictions, we get a good, more than qualitative agreement
with ``less short-distance'' inclusive decays as well. 
\vspace*{2.5mm}

A dedicated discussion of what may be wrong in a number of analyses
claimed problems for the OPE goes far beyond the scope of the present
talk. Two elements must be mentioned largely responsible
for this, they have been emphasized early enough \cite{uses,amst}, yet
were mostly ignored. One is imposing unnecessary and dangerous
constraint on the mass difference $m_b\!-\!m_c$ which relies on the
too questionable $1/m_c$ expansion, and applying the corresponding
power counting rules which are inadequate numerically. The
second is blind usage of the naive OPE expressions where their validity
is severely limited by barely sufficient effective hardness. The
terms exponential in the latter -- yet invisible in the naive approach
-- blow up there. 
Rephrasing Ben Bradley, one may 
then think that the alleged problems originate
from ``Dealing in the expressions, not necessarily in (OPE) truths''.
I would also mention that using the proper Wilsonian version of the
OPE with running quark masses and nonperturbative parameters is
helpful in arriving at correct conclusions, and is probably indispensable for
reaching the ultimate numerical precision in extracting the
underlying parameters.

\subsection{Semileptonic decays with \,{\Large \boldmath $\tau\nu_\tau$}}

Experimental studies of the semileptonic decays $B\!\to\! X_c+\tau\nu_\tau$
are much more difficult, however even accurate measurements may be
feasible at future high-luminosity super--$B$-factories. Relevant to
physics discussed in Sect.~2, the $B\!\to\!D \,\tau\nu_\tau$ amplitude
does not vanish at zero recoil when $\vec q \!\to\! 0$ keeping the
contribution of $f_-$ (still suppressed by the $\tau$ mass). This
opens an interesting way to check the BPS relation (\ref{106}) and its
accuracy dictated by the theorem (\ref{116}). This would complement an
independent determination of $V_{cb}$ through the small-recoil
$B\!\to\!D$ decay rate with light leptons. 

Since the velocity range is rather limited for decays into $\tau$
leptons, $1\!\le\! w\!\le\! 1.43\;$ the total ${\rm BR}(B\!\to\!D
\,\tau\nu_\tau)$ can be predicted in terms of $\varrho^2\!-\!\frac{3}{4}$,
at least in the BPS regime.

Inclusive decay width $\Gamma(B\!\to\! X_c+\tau\nu_\tau)$ is very
sensitive to $m_b\!-\!m_c$, in contrast to $m_b\!-\!0.6m_c$ for
conventional decays (the same applies to decay distributions), which
would finally provide a constraint on an alternative mass
combinations. 

On the other hand, energy release is limited here,
$m_b\!-\!m_c\!-\!m_\tau \!\simeq\!1.6\GeV$. These decays therefore can
be more sensitive to higher power corrections and represent a better
candidate for studying local duality violations. One can also detect
possible effects of the nonperturbative four-quark operators with
charm quark, $\matel{B}{\bar{b} c\,\bar{c}b}{B}$; their potential
role was emphasized in Ref.~\cite{imprec}.

I am curious if it is possible to measure here separately the decay
widths induced by vector and axial-vector currents. As emphasized in
Sect.~2, this makes a big difference for the nonperturbative effects in
the exclusive transitions to $D$ and $D^*$, respectively. Yet for
inclusive decays there should be no much difference once one enters the
short-distance regime. Therefore, such analysis would provide
interesting insights into the difference between the heavy quark expansion
for inclusive and exclusive decays.

\section{Conclusions} 

The theory of heavy quark decays is now a mature branch of
QCD. Recent experimental studies of inclusive decays yielded valuable
information crucial -- through the comprehensive application of all
the 
elements of the  heavy quark expansion -- for a number of exclusive 
decays as well. This signifies an important new stage in the heavy quark
theory, since only a few years ago exclusive and inclusive
decays were often viewed as largely separated, if not as antipodes 
theory-wise. 

Generally speaking, there is ample evidence that heavy quark symmetry
undergoes significant nonperturbative corrections for charm
hadrons. However, there appears a class of practically relevant
relations which remain robust. They are limited to the ground-state
pseudoscalar $B$ and $D$ mesons, but do not include spin symmetry in
the charm sector.

The accuracy of these new relations based on the proximity to the
``BPS limit'' strongly depends on the actual size of the kinetic
expectation value in $B$ mesons, $\mu_\pi^2(1\GeV)$. The experiment
must verify it with maximal possible accuracy and reliability, without
invoking ad hoc assumptions often made in the past with limited data
available. This can be performed through the inclusive decays already
in current experiments. If its value is confirmed not to 
exceed $0.45\GeV^2$, the $B\!\to\!D$ decays can be reliably treated by
theory, and the estimate ${\cal F}_+(0)\simeq 1.03$ can provide a good
alternative element in the comprehensive program of model-independent
extraction of $|V_{cb}|$. 

There are many other important consequences of the BPS regime. The
slope of the IW function must be close to unity, and actually below
it. The related constraints on the slope $\hat\rho^2$ of the 
experimentally observed combination of $B\!\to\!D^*$ formfactors
should be incorporated in the fits aiming at extrapolating the rate to
zero recoil. 

Contrary to what is often stated, the OPE-based theory of inclusive
decays so far is in good shape when confronted experiment. A nontrivial
consistency between quite different measurements, and between
experiment and QCD-based theory, at the nonperturbative level, has
been observed.
Yet this refers to a thoughtful application of the OPE using the robust
approach instead of doubtful approximations. There are good reasons
to hope that new round of data will not reverse the tendency, and that 
the comprehensive approach shall indeed allow us to reach a percent
level of reliable accuracy in translating $\Gamma_{\rm sl}(B)$ to
$|V_{cb}|$, as was proposed last year.

A final note -- I think that the semileptonic decays with $\tau$
lepton and $\nu_\tau$ have an interesting potential for both inclusive
and exclusive decays at high-luminosity machines.

\section*{Acknowledgments}

I would like to thank the organizers of FPCP\,2003 for the interesting
conference and for provided support. I am indebted to I.~Bigi and
P.~Gambino for joint work reported in this talk, and to M.~Shifman and
A.~Vainshtein for valuable discussions of the BPS regime. Many
discussions with and useful comments from O.~Buchmueller and
U.~Langenegger are gratefully acknowledged, which initiated parts of
the reported studies. This work was supported in part by the NSF under
grant number PHY-0087419.

\end{document}